\documentstyle[prl,aps,epsf]{revtex}

\begin{document}

\twocolumn[\hsize\textwidth\columnwidth\hsize\csname @twocolumnfalse\endcsname

\title{Quantum Monte Carlo simulations of the half-filled 
two-dimensional Kondo lattice model.}

\author{F.F. Assaad \\
   Institut f\"ur Theoretische Physik III, \\
   Universit\"at Stuttgart, Pfaffenwaldring 57, D-70550 Stuttgart, Germany. }

\maketitle

\begin{abstract}
The 2D half-filled Kondo lattice model with exchange $J$  
and nearest neighbor hopping $t$ is considered.  It is shown that this
model belongs to a class of Hamiltonians for which 
zero-temperature auxiliary field
Monte Carlo methods may be efficiently applied. 
We compute the staggered moment, spin and quasiparticle gaps on
lattice sizes up to $12 \times 12$. The competition between the 
RKKY interaction and Kondo effect leads to a continuous quantum 
phase transition between antiferromagnetic and spin-gaped insulators.
This transition occurs at $J/t = 1.45 \pm 0.05$. 
\end{abstract}
]

The Kondo lattice model (KLM) describes a band of conduction electrons 
interacting with local moments via an exchange interaction $J$.
This model is relevant for the understanding of heavy electron materials
\cite{Lee86,Aeppli92}. 
The nature of the ground state results from competing 
effects. 
The polarization cloud of conduction electrons produced 
by a local moment  may be felt by another local moment.  This 
provides the mechanism for the Ruderman-Kittel-Kasuya-Yosida (RKKY) 
interaction \cite{Kittel63}  
with effective exchange
$J_{eff}(\vec{q})  \propto  -J^2 {\rm Re} \chi (\vec{q}, \omega = 0) $,
$ \chi( \vec{q},\omega)$ being the spin  susceptibility of the 
conduction electrons. 
On the other hand,  the same polarization cloud may form a 
singlet bound state with the local moment. In the single impurity case, 
this happens at the Kondo temperature
$T_K  \propto \epsilon_f e^{-1/JN(\epsilon_f)}$ where $\epsilon_f$ is the
Fermi energy and $N(\epsilon_f)$ the density of states \cite{Yosida66}. 
Comparing energy scales, the RKKY interaction dominates at {\it small} $J$
and Kondo effect at {\it large} $J$. Thus a quantum 
transition between magnetically ordered and 
disordered  phases  is  anticipated.

The KLM we consider is  written as 
\begin{equation}
H_{KLM} = 
 -t \sum_{\langle \vec{i},\vec{j} \rangle ,\sigma } 
 c^{\dagger}_{\vec{i},\sigma} c_{\vec{j},\sigma} 
    + J \sum_{\vec{i}} 
    \vec{S}^{c}_{\vec{i}} \vec{S}^{f}_{\vec{i}}.
\end{equation}
Here $\vec{i}$ runs over the $L^2$-sites of a square lattice,
$\langle \vec{i},\vec{j} \rangle$ corresponds to nearest neighbors,
$c^{\dagger}_{\vec{i},\sigma} $ 
creates a conduction electron with 
z-component of spin $\sigma$  on site $\vec{i}$ and  periodic 
boundary conditions are imposed. 
$ \vec{S}^{f}_{\vec{i}} =(1/2) \sum_{\sigma,\sigma'} 
f^{\dagger}_{\vec{i},\sigma}    \vec{\sigma}_{\sigma,\sigma'} 
f_{\vec{i},\sigma'} $
and 
$ \vec{S}^{c}_{\vec{i}} =(1/2) \sum_{\sigma,\sigma'}
c^{\dagger}_{\vec{i},\sigma}    \vec{\sigma}_{\sigma,\sigma'}
c_{\vec{i},\sigma'} $ with 
$\vec{\sigma}$ the Pauli matrices.  
A constraint of one fermion per $f$-site is enforced. 
At $J/t \ll 1$ this model maps onto the periodic Anderson 
model \cite{Anderson61} (PAM) at strong coupling \cite{Schrieffer66}. 
Quantum Monte Carlo (QMC) \cite{White89}  
methods constitute an 
efficient tool for the study of the PAM  in various dimensions
\cite{Vekic95,Groeber98,Huscroft99}.
The one-dimensional version of  the KLM  has been extensively 
studied  \cite{Tsunetsugu97_rev}.
In particular, at $ \langle n \rangle = 2 $
(half-band filling or one conduction electron per local moment) 
the Kondo effect dominates at all values of $J/t$.
In two dimensions, variational Monte Carlo  methods \cite{Wang94}
as well as series expansions  around the strong coupling limit
\cite{Shi95} support the existence of a critical point.
The aim of this paper is to go beyond the above approximative
approaches. We  show how to efficiently simulate the
KLM with the projector 
QMC (PQMC) algorithm \cite{Koonin86,Sorella89}.   This method  yields 
{\it exact} zero temperature results and is free of  the notorious 
sign problem at half-band filling. This stands in contrast to previous
approaches \cite{Fye91}  which  generate a sign problem 
even at $\langle n \rangle = 2 $.
Using this algorithm, we map out the phase diagram
at $\langle n \rangle = 2 $.

Our starting point is the Hamiltonian:
\begin{eqnarray}
\label{H_QMC}
H = & &  -t \sum_{  \langle \vec{i},\vec{j} \rangle, \sigma }  
    c^{\dagger}_{\vec{i},\sigma} c_{\vec{j},\sigma}
    - \frac{J}{4} \sum_{\vec{i}} \left[   \sum_{\sigma}
    c^{\dagger}_{\vec{i},\sigma} f_{\vec{i},\sigma} + 
    f^{\dagger}_{\vec{i},\sigma} c_{\vec{i},\sigma}
   \right]^2  + \nonumber \\
	& & U_f \sum_{i} 
\left( n^{f}_{\vec{i},\uparrow} - 1/2 \right)
\left( n^{f}_{\vec{i},\downarrow}- 1/2 \right).
\end{eqnarray}
with $ n^{f}_{\vec{i},\sigma} = f^{\dagger}_{\vec{i},\sigma} 
f_{\vec{i},\sigma} $.
This Hamiltonian has all the properties required for an
efficient   use
of  QMC methods. 
To avoid working
with continuous fields, we use the approximate 
Hubbard Stratonovitch (HS)
transformation introduced in Ref. \cite{Assaad97,Motome97} to 
decouple the $J$-term.
This transformation introduces systematic errors of the
order  $(\Delta \tau)^3$, where  $\Delta \tau$   corresponds to  an 
imaginary time step. Since this order 
is higher than the systematic error produced by the Trotter decomposition,
it is negligible. As for the Hubbard term, we have found it essential 
to use Hirsch decomposition in terms of Ising spins which couple to
the density rather than to the $z$-component of the magnetization
\cite{Hirsch83}. 
Although  this  forces us to work with complex numbers, 
it conserves $SU(2)$-symmetry for a given HS configuration.  As argued 
in Ref. \cite{Assaad98b}, this provides an efficient algorithm for  
the calculation of imaginary time displaced spin-spin correlation 
functions  \cite{Assaad96a}
from which we will determine the spin gap.  
The ground state of $H$ (\ref{H_QMC}), $ | \Psi_0 \rangle $,
is obtained by projection. A trial wave function, 
$ | \psi_T\rangle$, required to be  a
single Slater determinant and non-orthogonal to the ground state is 
propagated along the imaginary time axis till convergence is reached
\cite{Koonin86,Sorella89}. 
With the above HS 
transformations and appropriate choice of $ | \Psi_T \rangle$
\cite{Assaad97},
particle-hole symmetry  leads to the absence of sign
problem at $\langle n \rangle = 2$. 

\begin{figure}
\epsfxsize=8.0cm
\hfil\epsfbox{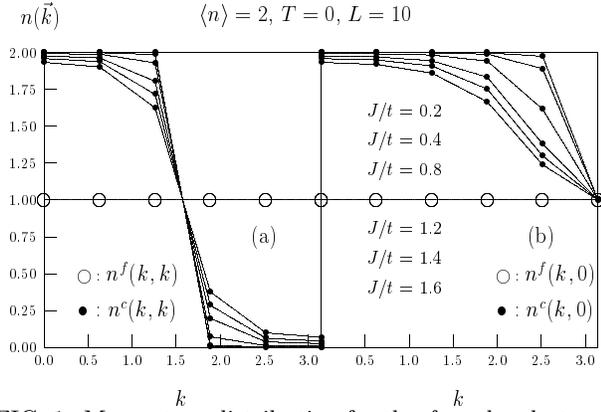}\hfil
\caption[]
{\noindent Momentum distribution for the $f$- and $c$-electrons.
(a)  $\vec{k} = k (1,1)$  and (b) $\vec{k} = k(1,0)$.
Within our precision, 
$n^{f} (\vec{k}) \equiv 1 $ for all 
considered values of $J/t$. To achieve this we have chosen 
values $U_f/t$ ranging from $U_f/t=0.5$ ($J/t =0.2$) to 
$U_f/t = 2$ ($J/t = 1.6$).  As $J/t$ grows, $n^{c}(\vec{k})$ 
becomes smoother. 
\label{nk.fig} }
\end{figure}

The relation of the above model to the KLM model is seen by  rewriting
Eq. \ref{H_QMC}  as
\begin{eqnarray}
H = & &  -t \sum_{  \langle \vec{i},\vec{j} \rangle, \sigma }  
    c^{\dagger}_{\vec{i},\sigma} c_{\vec{j},\sigma}
    + J \sum_{\vec{i}} 
    \vec{S}^{c}_{\vec{i}} \vec{S}^{f}_{\vec{i}}  -  \nonumber \\
     & & J \sum_{\vec{i}} 
   \left( c^{\dagger}_{\vec{i},\sigma} c^{\dagger}_{\vec{i},-\sigma}
         f_{\vec{i},-\sigma} f_{\vec{i},\sigma}  + {\rm H. c.}
   \right)  + \nonumber \\
    & &  J \sum_{\vec{i}} \left( n^{c}_{\vec{i}} n^{f}_{\vec{i}} -
                 n^{c}_{\vec{i}} - n^{f}_{\vec{i}} \right)
	+  \nonumber \\
   & & U_f \sum_{i}  
   \left( n^{f}_{\vec{i},\uparrow} - 1/2 \right)
   \left( n^{f}_{\vec{i},\downarrow}- 1/2 \right).
\end{eqnarray}
with $ n^{f}_{\vec{i}} = \sum_{\sigma} f^{\dagger}_{\vec{i},\sigma} 
f_{\vec{i},\sigma} $  and 
$ n^{c}_{\vec{i}} = \sum_{\sigma} c^{\dagger}_{\vec{i},\sigma} 
c_{\vec{i},\sigma} $.
It is important to notice that 
\begin{equation}
 [H,  \sum_{\vec{i}}
    (1 - n^{f}_{\vec{i},\uparrow}) (1 - n^{f}_{\vec{i},\downarrow}) 
    + n^{f}_{\vec{i},\uparrow} n^{f}_{\vec{i},\downarrow} ] = 0.
\end{equation}
The number of doubly occupied and empty $f$-sites is a conserved quantity.
Denoting by $P_n$ the projection onto the subspace with $n$ doubly 
occupied and empty
$f$-sites  one obtains:
\begin{equation}
\label{HP0}
      H P_0 =   H_{KLM} -  N(U_f/4 + J).
\end{equation}
Thus, in principle, it suffices to consider  a trial wave function satisfying
$P_0 | \Psi_T \rangle  = | \Psi_T \rangle  $ to ensure that 
$ \exp(- \Theta H) | \Psi_T \rangle  = \exp(- \Theta H_{KLM}) | \Psi_T 
\rangle $. 
The coupled constraints: 
$P_0 | \Psi_T \rangle  = | \Psi_T \rangle  $   and  
$| \Psi_T  \rangle  $ is a Slater determinant forces us to  choose
$S_{\vec{i}}^{f,z} | \Psi_T \rangle = \pm \frac{1}{2} | \Psi_T \rangle $ 
thus breaking $SU(2)$-spin symmetry.
Since the KLM  conserves total spin, 
this symmetry has to be restored by the imaginary time 
propagation. 
When the energy gap to
the first excited spin state is small - as is certainly
the case when long-range magnetic order is present-
restoring this symmetry is extremely expensive.
To avoid this  problem and 
since the ground state of the KLM at half-filling on a bipartite lattice 
has $S=0$ \cite{ShenSQ96,Tsunetsugu97a}, 
we choose a spin singlet trial wave function.
During the 
imaginary time propagation
$P_n | \Psi_T \rangle $ will be suppressed by a factor 
$e^{-n \Delta E \Theta }$ in comparison to $ P_0 | \Psi_T \rangle $.  
In two limiting cases, we estimate
$\Delta E  \sim U_f/4$  for $J/t \ll 1  $ 
and $ \Delta E  \sim  3 U_f/8$ for  $ J/t, J/U_f \gg 1$. 
To confirm that we are well in the $P_0$ subspace we 
plot in Fig. 
\ref{nk.fig}
the single particle occupation number
$ n_{\vec{k}}^{f} \equiv 
\langle  \Psi_0  | \sum_{\sigma}  f^{\dagger}_{\vec{k},\sigma} 
f_{\vec{k}, \sigma}  | \Psi_0 \rangle $ with 
$ f_{\vec{k}, \sigma}  = (1/L) \sum_{\vec{j}} e^{i \vec{k} \cdot \vec{j} }
f_{\vec{j},  \sigma} $.
Our results are 
indistinguishable from $ n_{\vec{k}}^{f} \equiv 1 $ which leads to
$ \langle  \Psi_0  | \sum_{\sigma}  f^{\dagger}_{\vec{i},\sigma}
f_{\vec{j}, \sigma}  | \Psi_0 \rangle  = \delta_{\vec{i},\vec{j}}$, 
a property which may be only realized if $P_0 |\Psi_0 \rangle  =
|\Psi_0 \rangle$.  Owing to  Eq. (\ref{HP0}) $ | \Psi_0 \rangle $ 
is nothing but the ground state of the KLM.

\begin{figure}
\epsfxsize=8.0cm
\hfil\epsfbox{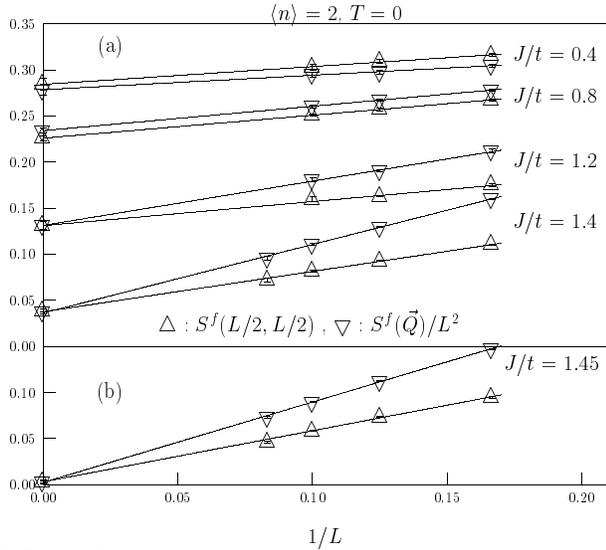}\hfil
\caption[]
{\noindent  Spin-spin correlations for the $f$-electrons versus inverse
linear length $L$ for several values of $J/t$.  The solid lines are 
least square fits to the form $a + b/L$. The symbol at $1/L = 0$ corresponds
to the extrapolated value. As apparent, both $S^{f}(L/2,L/2)$
and $S^{f}( \vec{Q} )/L^2$ scale to the same value.
\label{spinff.fig} }
\end{figure}

\begin{figure}
\epsfxsize=8.0cm
\hfil\epsfbox{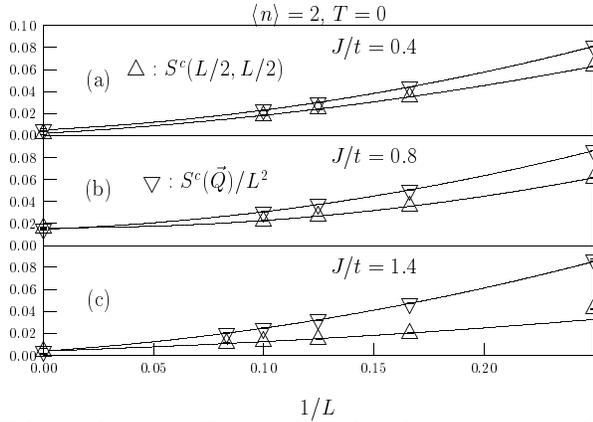}\hfil
\caption[]
{\noindent
Same as Fig. \ref{spinff.fig} but for the conduction electrons. In order 
to satisfy the  relation 
$ \lim_{L \rightarrow \infty } S^{f}(L/2,L/2)$  $\equiv  $
$ \lim_{L \rightarrow \infty } S^{f}( \vec{Q} )/L^2$  we fit the data 
to the form  $a + b/L + c/L^2$.
\label{spinss.fig} }
\end{figure}

\begin{figure}
\epsfxsize=8.0cm
\hfil\epsfbox{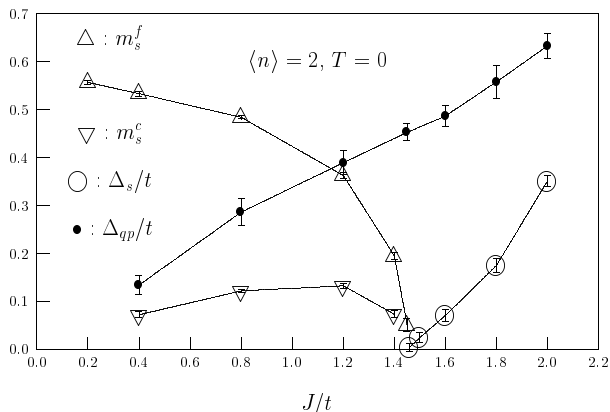}\hfil
\caption[]
{\noindent  Phase diagram of the KLM at half-band filling and $T=0$.  All
plotted quantities are extrapolated to the thermodynamic limit. 
At $J/t = 0.2$ we were not able to distinguish $m^c_{s}$ from zero.
\label{phase.fig} }
\end{figure}

We now discuss  the  phase diagram at $\langle n \rangle = 2 $
and start with the spin degrees of freedom. 
To establish long-range magnetic order, we compute the quantities 
$ S^{\alpha}(\vec{r}) =
\frac{4}{3} \langle \vec{S}^{\alpha}(\vec{r}) \cdot \vec{S}^{\alpha}(\vec{0}) 
\rangle  $ as well as its  Fourier transform:
$ S^{\alpha}(\vec{q}) = \sum_{\vec{r}} e^{i \vec{q} \cdot \vec{r} } 
S^{\alpha}(\vec{r})$.  We  consider separately the conduction 
($\alpha = c$) and  localized ($\alpha = f$) electrons.  
Long-range antiferromagnetic order is present  when
$ \lim_{ L \rightarrow  \infty } S^{\alpha}(L/2,L/2)   \equiv 
  \lim_{L  \rightarrow \infty } S^{\alpha}(\vec{Q} \equiv (\pi,\pi))/L^2 $ 
takes a finite value.
Fig. \ref{spinff.fig}  plots both  above quantities 
versus  $1/L$ for the $f$-electrons. 
For lattice sizes ranging from $L=6$ to $L=12$ the 
QMC data extrapolates linearly to a finite value for $J/t \leq 1.45$.
Similar results are plotted in  Fig. \ref{spinss.fig} for the 
conduction electrons. 
The resulting  staggered moment 
$ m_s^{\alpha} \equiv  
\sqrt{  \lim_{L\rightarrow \infty} S^{\alpha}(\vec{Q})/L^2} $ 
is plotted versus $J/t$ in Fig. \ref{phase.fig}.
At $J/t = 0.2 $,  $m_s^{f} = 0.557(3)$ - a value much larger
than for the Heisenberg model: $m_s^H = 0.3551(3)$ \cite{Sandvik97}. 
In contrast,
$m_s^{c}$ is  small  at small values of $J/t$ -
$m_s^{c} = 0.072(6) $  at J/t = 0.4. In comparison, the half-filled 
Hubbard model at $U/t =4$ leads to   $m_s  \sim 0.2 $ \cite{White89}. 
\noindent
At the mean-field level, the  behavior of  $ m_s^{f,c}$ at weak
coupling may be captured  by the   Ansatz,  
$\langle \vec{S}^{f}_{\vec{i}} \rangle = (-1)^{\vec{i}}
\tilde m^{f}_s \vec{e}_z/2 $ ($ \tilde m^{f}_s  \leq 1$)  which  
leads  to $\tilde H_{KLM} = 
-t \sum_{\langle \vec{i}, \vec{j} \rangle, \sigma} 
c^{\dagger}_{\vec{i},\sigma} c_{\vec{j},\sigma}
+ (J \tilde m^{f}_s /2) \sum_{\vec{i}} \vec  (-1)^{\vec{i}} 
{S}^{c}_{\vec{i}} \cdot 
\vec{e}_z $. Minimizing the free energy  with respect to 
$\tilde m^{f}_s$ yields:  $\tilde m^{f}_s  = 1$.
The conduction electrons are thus subject to a staggered field of
magnitude $ \propto  J$.  Since  ${\rm Re} \chi ( \vec{Q}, \omega = 0  ) $
is singular,
this immediately leads to long-range magnetic order with
$\tilde m_s^{c}  \propto  (J/t) \ln^2(J/t)$ for $J/t  \ll 1$. 
The behavior of $\tilde m_s^{c,f} $ bears some similarity with the
QMC data  (see Fig. \ref{phase.fig}).
At larger values of $J/t$ the Kondo effect  destroys magnetic order.
Both $m_s^f$ and $m_s^c$ scale {\it continuously} to zero as $J/t $ approaches
$J_c/t \sim 1.45$  (see Fig. \ref{phase.fig}).

\begin{figure}
\epsfxsize=8.0cm
\hfil\epsfbox{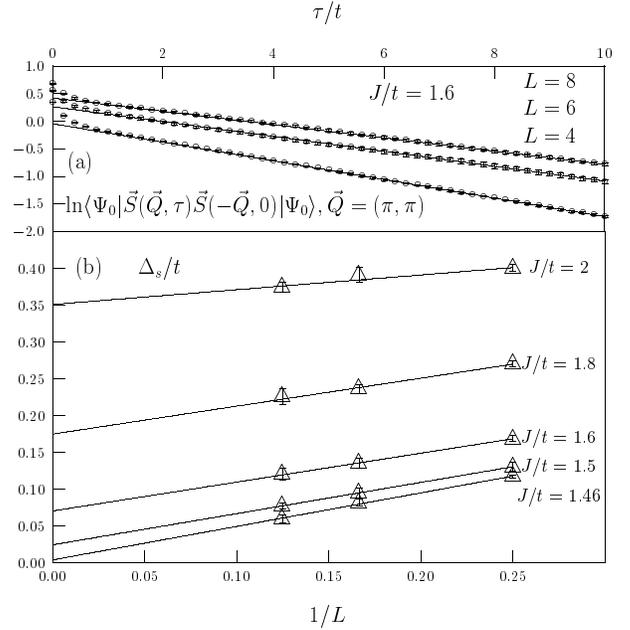}\hfil
\caption[]
{\noindent 
(a) $ \ln S(\vec{Q},\tau) $ versus $\tau t$ for $J/t = 1.6$.  
The solid lines correspond to least square  of the tail of 
$ S(\vec{Q},\tau) $ to the form  a $e^{-\tau \Delta_s}$. The thus 
obtained value of the spin gap $\Delta_s$ is plotted versus $1/L$ in (b). 
The solid line in (b) are least square fits to the form $a + b/L$.
\label{spgap.fig} }
\end{figure}

Once long-range magnetic order is destroyed
($J/t > 1.45$) the ground state is expected to evolve smoothly to the
strong coupling limit, $J/t  \gg 1$. In this  limit, $| \Psi_0 \rangle$  is
given by a direct product of singlets on the $f$-$c$ bonds
of an elementary cell.  Starting from this state, a triplet excitation 
has a dispersion relation  (up to second order in $t/J$) 
$\Delta_s^{(2)}(\vec{q}) 
= J - \frac{16t^2}{3J} - \frac{2t}{J} \epsilon(\vec{q})$
\cite{Tsunetsugu97_rev}. 
To compute $\Delta_s(\vec{q})$  numerically we consider $S(\vec{q},\tau) = $
$ \frac{4}{3} \langle \Psi_0 | \vec{S}(\vec{q},\tau) \cdot \vec{S}(-\vec{q},0)
| \Psi_0 \rangle$ where 
$ \vec{S}(\vec{q}) \equiv \vec{S}^{f}(\vec{q}) + \vec{S}^{c}(\vec{q}) $
and $ \vec{S}(\vec{q},\tau) = e^{\tau H } \vec{S}(\vec{q}) e^{-\tau H }$.
For $\tau  t \gg 1$,  $S(\vec{q},\tau) \propto 
\exp \left( -\tau  \Delta_s(q) \right)$ with
$\Delta_s (\vec{q}) \equiv E_0(S=1,N,\vec{q}) - E_0(S=0,N)$. Here
$E_0(S,N,\vec{q})$ denotes the ground state energy with total spin $S$,
momentum $\vec{q}$ and particle number $N \equiv L^2$. 
As in the strong coupling limit 
and for all considered values
of $J/t$  the spin gap 
$\Delta_s \equiv {\rm min}_{\vec{q}} \Delta_s(\vec{q}) = \Delta_s(\vec{Q})$
with $\vec{Q} = (\pi,\pi)$.
Fig. \ref{spgap.fig}a plots the raw data  from which we obtain the spin-gap
and  Fig. \ref{spgap.fig}b,  $\Delta_s$ versus $1/L$.
A linear
extrapolation to the thermodynamic limit leads to the  results plotted in
Fig. \ref{phase.fig}. 
Within our accuracy, the value of $J/t$ for which long-range
magnetic order vanishes corresponds to the value of $J/t$ where  the 
spin-gap vanishes.

\begin{figure}
\epsfxsize=8.0cm
\hfil\epsfbox{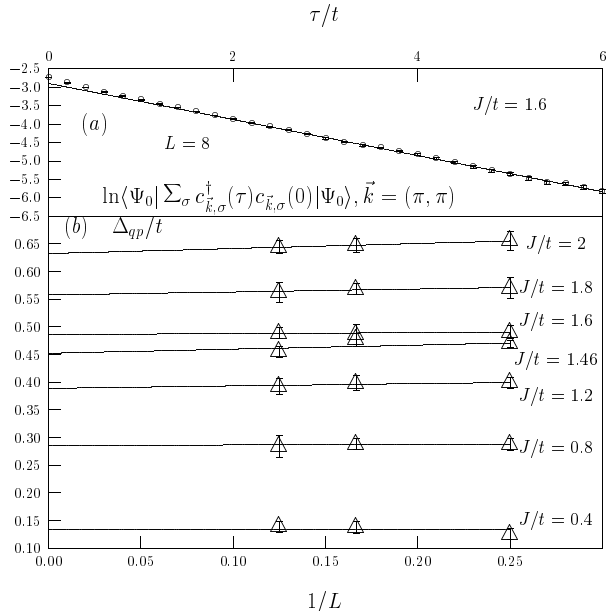}\hfil
\caption[]
{\noindent
Same a Fig. \ref{spgap.fig} but for the quantity
$\sum_{\sigma}
\langle \Psi_0 | c^{\dagger}_{\vec{k},\sigma }(\tau) c_{\vec{k},\sigma }
|  \Psi_0 \rangle $ so as to obtain $\Delta_{qp}$.
\label{qugap.fig} }
\end{figure}

Finally, we consider the quasiparticle gap.   As apparent from the
single particle occupation number, $ n^{c}(\vec{k})$, (Fig. \ref{nk.fig})
the quasiparticle gap grows continuously with growing values of $J/t$.  
To obtain an accurate
estimate  of this quantity, we compute 
$ \langle \Psi_0 | \sum_{\sigma} 
c^{\dagger}_{\vec{k},\sigma }(\tau) c_{\vec{k},\sigma }
|  \Psi_0 \rangle $   which scales as 
$e^{-\tau \Delta_{qp}(\vec{k})} $  when $\tau  t \gg 1$.
Here, $\Delta_{qp} (\vec{k}) = E_0(N) - E_0(N-1,\vec{k})$.
In the strong coupling limit and to  first order in $t/J$,  
$\Delta^{(1)}_{qp} (\vec{k}) =  3J/4 - \epsilon(\vec{k})/2 $
and thus takes a minimum at $\vec{k} = (\pi,\pi)$. 
Our QMC results for values of  $J/t$ ranging
from $J/t=0.4$ to $J/t = 2$ are consistent with:
$\Delta_{qp} \equiv {\rm min}_{\vec{k}} \Delta_{qp}(\vec{k})  = 
\Delta_{qp}(\pi,\pi)$  \cite{Note1}.
The size scaling of $ \Delta_{qp} $
is presented in Fig. \ref{qugap.fig} and the extrapolated value 
is plotted in Fig. \ref{phase.fig} versus $J/t$. As apparent,
$\Delta_{qp} $  remains finite and  evolves {\it smoothly }  through 
the quantum transition. In the above discussed mean-field approach 
based on the Ansatz
$\langle \vec{S}^{f}_{\vec{i}} \rangle = (-1)^{\vec{i}}
\tilde m^{f}_s \vec{e}_z/2 $,  the 
quasiparticle gap scales as $J/4$  in the small $J/t$ limit. Such a
behavior, equally seen in one-dimension \cite{Tsunetsugu97_rev,Yu93}, is to a 
first approximation consistent with our data
(see Fig. \ref{phase.fig}).

To summarize, we have presented an efficient auxiliary field 
QMC algorithm to simulate zero-temperature properties of the KLM.
At half-band filling where the sign-problem is absent we 
calculated the staggered moment, the spin gap, and the 
quasiparticle gap on lattice sizes up to $12  \times 12 $.  Our results
are summarized in Fig. \ref{phase.fig}. We observe a
{\it continuous} quantum phase transition between long-range 
antiferromagnetic and
spin-gaped phases. This transition occurs at $J_c/t = 1.45 \pm 0.05$ 
in good agreement with previous approximative results \cite{Wang94,Shi95}.
The quasiparticle gap is finite and evolves {\it continuously}
between both phases. Given that the  charge degrees 
of freedom remain gaped, we expect the observed quantum phase 
transition to belong to the universality class of the $O(3)$ 
nonlinear sigma model \cite{Chakravarty89,Chubukov94}.

A. Muramatsu is thanked for instructive conversations. 
The simulations were carried out on the T3E of the HLRS-Stuttgart, as
well as on the T90 and T3E of the HLRZ-J\"ulich.

\end{document}